\documentclass[aps,pra,twocolumn,superscriptaddress,longbibliography]{revtex4-2}

\usepackage{graphicx}
\usepackage{dcolumn} 
\usepackage{bm}      
\usepackage{amsmath}
\usepackage{amssymb}
\usepackage{epstopdf}
\usepackage{xcolor}
\usepackage{comment}
\usepackage[normalem]{ulem}
\usepackage{hyperref}

\usepackage[normalem]{ulem}

\begin{document}
\title{Intercavity phonons and dynamics in coupled polariton cavities}
\author{Iliana Carmona-Moreno}
\author{Grover Andrade-S\'anchez}
\author{Hugo A Lara-García}
\author{Giuseppe Pirruccio}
\author{Arturo Camacho-Guardian}
\email{acamacho@fisica.unam.mx}
\affiliation{Departamento de F\'isica Qu\'imica, Instituto de F\'isica, Universidad Nacional Aut\'onoma de M\'exico, Apartado Postal 20-364, Ciudad de M\'exico C.P. 01000, Mexico\looseness=-1}
\date{\today}
\begin{abstract}
Intercavity polaritons, hybrid quasiparticles with spatially separated photonic and excitonic components, provide a platform to engineer structured light–matter states. We show that resonant driving of the middle polariton branch leads to a qualitatively distinct dynamical regime in which coherent Rabi oscillations are suppressed, and the system evolves monotonically toward its steady state. Including interactions, we demonstrate that this regime supports Bogoliubov excitations with a phonon-like dispersion at low momenta. These collective modes inherit interactions from the excitonic fraction, while preserving the intrinsically intercavity nature of the quasiparticles.

\end{abstract}

\maketitle

\section{Introduction}

Polaritons are quasiparticles that arise from the strong coupling between cavity photons and matter excitations~\cite{Hopfield1958}, combining the effective light mass of photons with the matter interactions of excitons. This  hybrid character has enabled the realization of a wide range of collective phenomena in solid-state platforms, including room-temperature Bose-Einstein condensation~\cite{plumhof2014room,daskalakis2014nonlinear,Carusotto2013}, superfluid flow~\cite{lerario2017room}, quantum vortices~\cite{lagoudakis2008quantized} and persistent currents~\cite{sanvitto2010persistent,sanvitto2011all,keeling2011superfluid}, as well as strongly interacting regimes such as polariton Feshbach resonances~\cite{takemura2014polaritonic,sidler2017fermi}. Beyond their fundamental interest, these effects establish polaritons as a versatile platform for exploring nonequilibrium many-body physics and engineering quantum photonic devices~\cite{kavokin2022polariton,marsault2015realization}.  

A central advance in this direction has been the engineering of \textit{polariton lattices}, where spatially patterned potentials tailor the polariton dispersion. By imposing external modulation through lithographic strain engineering, or all-optical potentials, it is possible to design polariton lattices~\cite{amo2016exciton,su2020observation,Smirnova2024,Whittaker2018}, topological band structures~\cite{klembt2018exciton}, and realize polariton condensation in lattices~\cite{Winkler_2015,dusel2020room,Klembt2017,Suchomel2018,dusel2020room} thus emulating paradigms of condensed-matter physics in driven-dissipative settings~\cite{berloff2017realizing}. These systems provide a versatile testbed for quantum simulation, enabling on-demand control of band structures and localization effects, and thus expanding the scope of polaritonics toward scalable photonic circuitry.    

Strongly coupled cavity architectures provide a powerful route to engineer light–matter states with enhanced control. In particular, intercavity polaritons are hybrid quasiparticles whose photonic and excitonic components are spatially separated across distinct cavities while remaining coherently coupled. This spatial segregation enables independent tuning of light and matter degrees of freedom while preserving strong coupling. Intercavity polaritons have been experimentally realized in organic semiconductor platforms~\cite{GarcaJomaso2024}, establishing a simple and scalable architecture. Notably, this approach allows for the emergence of flat bands~\cite{sanchez2025protecting} and controlled mode hybridization, opening new avenues for quantum state engineering, tomography, and manipulation.

\begin{figure}[h]
    \centering
    \includegraphics[width=0.5\textwidth]{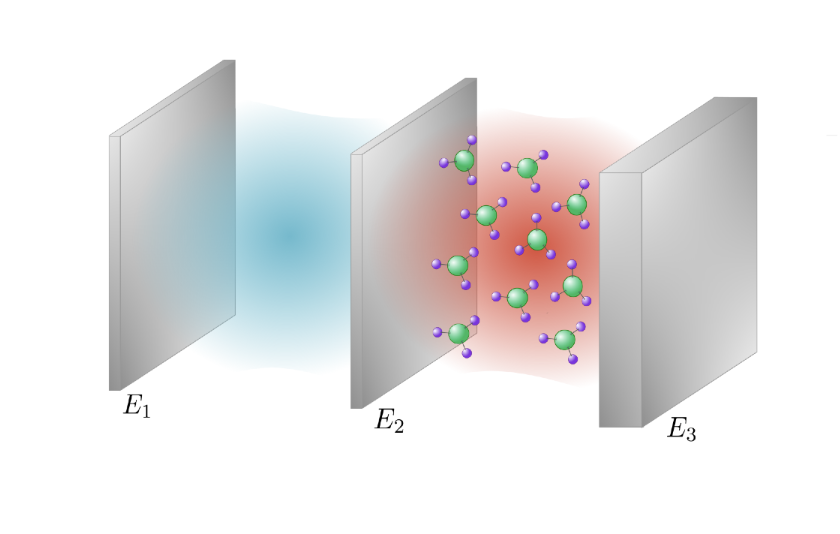}
    \caption{Schematic representation of two strongly coupled cavities. The left cavity contains only a dielectric medium, while the right cavity hosts an excitonic resonance. The cavities are coupled via a thin metallic mirror that mediates photon tunneling.}
    \label{fig:cartoon}
\end{figure}

So far, most theoretical studies of intercavity polaritons have focused on their steady-state properties and have largely relied on equilibrium-inspired descriptions. However, the intrinsically driven–dissipative character of polaritons can qualitatively modify their behavior, giving rise to nontrivial dynamical regimes far from equilibrium. Understanding these effects therefore requires going beyond static considerations to explicitly address their dynamics. In this context, intercavity architectures offer a unique opportunity for engineering quasiparticles and collective excitations with spatially segregated photonic and excitonic components, enabling protected hybrid states that retain a heavy mass component of the photons while inheriting interactions from matter degrees of freedom. 

In this work, we investigate the dynamical formation and collective excitations of intercavity polaritons in a driven–dissipative setting, establishing a direct connection between nonequilibrium dynamics and emergent many-body behavior. We show that resonant excitation of the middle polariton branch leads to a qualitatively distinct dynamical regime, where the system evolves monotonically toward a steady state with a strongly suppressed right-cavity photonic component, in contrast to the Rabi oscillations observed for the lower and upper branches. This behavior originates from the spatial segregation of the hybridized modes, which  suppresses  coherent Rabi exchange. Building on this dynamical steady state, we then incorporate exciton–exciton interactions and demonstrate that the middle branch supports a macroscopically occupied state with well-defined Bogoliubov excitations. These exhibit a phonon-like dispersion at low momenta, inheriting interactions from the excitonic fraction while retaining the intercavity character of the underlying state. Our results thus provide a unified framework linking nonequilibrium formation dynamics with collective excitations in spatially structured polaritonic systems.

\section{Model}
\label{Model}
We consider a system formed by two strongly coupled cavities, schematically shown in Fig.~\ref{fig:cartoon}. The left cavity is filled with a polymer but does not host an active medium, while the right cavity contains an organic semiconductor supporting an excitonic resonance. The two cavities are coupled via a thin metallic mirror, which enables photon tunneling between them.

In equilibrium, the system is described by the Hamiltonian
\begin{gather}
\hat H_0=\omega_L \hat a_L^\dagger \hat a_L+\omega_R \hat a_R^\dagger \hat a_R+\omega_X \hat x^\dagger \hat x \nonumber \\
 -J\big(\hat a_L^\dagger \hat a_R+\hat a_R^\dagger \hat a_L\big) 
 +\Omega\big(\hat a_R^\dagger \hat x+\hat x^\dagger \hat a_R\big),
\end{gather}
where $\hat a_{L/R}^\dagger$ creates a photon in the left/right cavity with energy $\omega_{L/R}$, and $\hat x^\dagger$ creates an exciton in the right cavity with energy $\omega_X$. The parameter $J$ denotes the photon tunneling amplitude between the cavities, while $\Omega$ is the Rabi coupling between excitons and right-cavity photons.  

\paragraph*{Polariton branches and Hopfield decomposition.}
Diagonalizing $\hat H_0$ in the bare basis $\{|L\rangle\equiv \hat a_L^\dagger|0\rangle,\ |R\rangle\equiv \hat a_R^\dagger|0\rangle,\ |X\rangle\equiv \hat x^\dagger|0\rangle\}$ yields three hybrid eigenmodes $|j\rangle$ with $j\in\{\mathrm{LP},\mathrm{MP},\mathrm{UP}\}$.
Each mode can be written as a coherent superposition
\begin{gather}
|j\rangle = C_L^{(j)}|L\rangle + C_R^{(j)}|R\rangle + C_X^{(j)}|X\rangle, \qquad 
Z_\alpha^{(j)} \equiv |C_\alpha^{(j)}|^2,
\end{gather}
where $\alpha\in\{L,R,X\},$ and $Z_\alpha^{(j)}$ denote the (squared) Hopfield coefficients, which play a role analogous to quasiparticle residues in the sense of weighting the contribution of each bare component. These satisfy $Z_L^{(j)}+Z_R^{(j)}+Z_X^{(j)}=1.$

In general, the eigenmodes of $\hat H_0$ yield cumbersome expressions. However, under resonance conditions ($\omega_L=\omega_X$), the spectrum simplifies to three polariton branches with energies
\begin{gather}
\omega_{\text{UP}/\text{LP}}=\frac{\omega_L+\omega_R \pm \sqrt{(\omega_L-\omega_R)^2+4(\Omega^2+J^2)}}{2}, \\
\omega_{\text{MP}}=\omega_L.
\end{gather}
The middle polariton (MP) is of particular interest, as it appears as a purely intercavity mode. Its field operator takes the form
\begin{gather}
 \hat b_{\text{MP}}=\cos\theta\,\hat a_L 
 + \sin\theta\,\hat x,
\end{gather}
with $\tan \theta=J/\Omega,$ showing that the MP is a coherent superposition of left-cavity photons and right-cavity excitons, completely decoupled from right-cavity photons. In this case, the residues are $Z_L^{\text{(MP)}}=\cos^2\theta,$ $Z_R^{\text{(MP)}}=0$ and $Z_X^{\text{(MP)}}=\sin^2\theta.$

To probe the system, we introduce a coherent drive applied to the left cavity, described by
\begin{gather}
\hat H_{\text{drive}} = 
F\big(e^{-i\omega_p t}\hat a^\dagger_L+e^{i\omega_p t}\hat a_L\big),
\end{gather}
where $F$ and $\omega_p$ denote the pump amplitude and frequency, respectively. Dissipation is incorporated phenomenologically via complex energy shifts $\omega_\alpha \rightarrow \omega_\alpha - i\gamma_\alpha$ ($\alpha=L,R,X$), accounting for photon leakage and nonradiative exciton decay. This corresponds to an effective non-Hermitian description of the driven–dissipative dynamics.  

\section{Polariton Dynamics}
\label{polaritondynamics}
To analyze the out-of-equilibrium dynamics, we define the spinor
\begin{gather}
\hat \Psi = 
\big[\hat a_L(t),\;\hat a_R(t),\;\hat x(t)\big]^T,
\end{gather}
and write the Heisenberg equations of motion in the rotating frame of the pump frequency $\omega_p$ as
\begin{gather} \label{eq:3_Heisenberg_motion_rot_frame}
  i\frac{\partial \hat \Psi(t)}{\partial t}=\hat H \cdot \hat \Psi(t)+\mathbf F,
\end{gather}
with the driving vector $\mathbf F=[F,\,0,\,0]^T$, and effective Hamiltonian
\begin{gather} \label{eq:effective_hamiltonian}
\hat H=
\begin{bmatrix} 
 -\Delta_L & -J & 0 \\
 -J & -\Delta_R & \Omega \\
 0 & \Omega & -\Delta_X
\end{bmatrix},    
\end{gather}
where $\Delta_\alpha=\omega_p-\omega_\alpha+i\gamma_\alpha$.

We obtain an analytical expression for the steady-state solution. In the long-time limit $t\to\infty$, we find
\begin{widetext}
\begin{gather}
\langle\hat a_L(t\rightarrow\infty)\rangle=\sqrt{N_{\text{tot}}^{(0)}}\left[\frac{\Omega}{\sqrt{\Omega^2+J^2}\left(1-\frac{\Delta_X\Delta_R}{\sqrt{\Omega^2+J^2}}\right)}-\frac{\Delta_X\Delta_R}{\Omega\sqrt{\Omega^2+J^2}\left(1-\frac{\Delta_X\Delta_R}{\sqrt{\Omega^2+J^2}}\right)}\right]\approx \sqrt{N_{\text{tot}}^{(0)}}\cos\theta +\mathcal O(\Delta_X) \\
\langle\hat a_R(t\rightarrow\infty)\rangle=\sqrt{N_{\text{tot}}^{(0)}}\Delta_X\left[\frac{1}{\sqrt{\Omega^2+J^2}\left(1-\frac{\Delta_X\Delta_R}{\sqrt{\Omega^2+J^2}}\right)}\right]\approx \mathcal O(\Delta_X)\\
\langle\hat x(t\rightarrow\infty)\rangle=\sqrt{N_{\text{tot}}^{(0)}}\left[\frac{J}{\sqrt{\Omega^2+J^2}\left(1-\frac{\Delta_X\Delta_R}{\sqrt{\Omega^2+J^2}}\right)}\right]\approx \sqrt{N_{\text{tot}}^{(0)}}\sin\theta +\mathcal O(\Delta_X), 
\end{gather}
\end{widetext}
with $N_{\text{tot}}^{(0)}=\left|\frac{\Omega F}{\sqrt{\Omega^2+J^2}\Delta_X}\right|^2, $ see further details in the Appendix. The total number of excitations scales quadratically with $F$ and inversely with $\Delta_X$. The latter indeed controls the formation of {\it intercavity}  polaritons, that is, for $\Delta_X$ small, we have
\begin{gather}
N\approx \underbrace{N_{\text{tot}}^{(0)}\cos^2\theta}_{\text{left photons}}+\underbrace{N_{\text{tot}}^{(0)}\sin^2\theta}_{\text{excitons}},
\end{gather}
since the contribution of the right cavity photons scaling is linear with $\Delta_X,$ the right cavity photons are strongly suppressed for $\Delta_X$ small, in contrast to the occupation of the modes of the left cavity photons and excitons.
That is, under proper pump and detuning conditions, the excitations are a linear superposition of left cavity photons and excitons, with a negligible contribution of the right cavity photon, forming essentially an {\it intercavity polariton}. 

In our numerics, we focus on the experimentally relevant case  of Frenkel intercavity polaritons~\cite{GarcaJomaso2024}, and take parameters consistent with organic polaritons. These results can be straightforwardly extended to other organic, inorganic or hybrid semiconductors. We take the cavity energies of the order $\omega_L=2.24 \; \text{eV}$ and $\omega_R=2.35 \; \text{eV}$, while the exciton resonance is chosen as $\omega_X=2.24 \; \text{eV}$ \cite{GarcaJomaso2024, sanchez2025protecting}. 
We work in the strong light-matter regime and take $\Omega/\gamma_X=10.$ We  start by varying the ratio $J/\Omega$, which  can be realized experimentally by modifying the thickness of the internal mirror. We also vary the pump frequency to probe the different polariton branches as explained below. Throughout, the system is driven via the left cavity, as specified in the Hamiltonian.
  
\begin{figure*}[t]
    \centering
    \includegraphics[width=\textwidth]{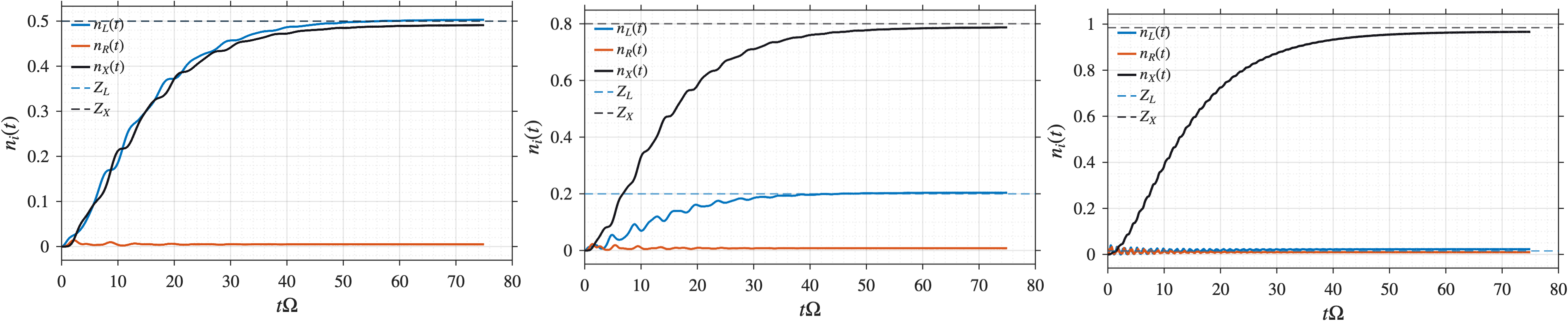}
\caption{\textbf{Dynamical formation of intercavity polaritons.} 
    Time evolution of the populations $n_L(t)$, $n_R(t)$, and $n_X(t)$ associated with the left cavity photon, right cavity photon, and exciton components, respectively, under resonant pumping of the middle polariton branch. Panels show three representative tunneling-to-coupling ratios: (a) $J/\Omega = 1$, (b) $J/\Omega = 2$, and (c) $J/\Omega = 4$. Dashed horizontal lines indicate the asymptotic steady-state values $Z_L$ and $Z_X$ predicted from the polariton eigenmode composition.}
    \label{fig:D1}
\end{figure*}

In Fig.~\ref{fig:D1} we show the formation of the intercavity polariton for several values of the ratio $J/\Omega$. The solid lines depict the normalised occupation of the left cavity photons (blue), the right cavity photons (red) and the excitons (black). The dashed lines following the same color coding depict the equilibrium Hopfield coefficients $Z_L$ and $Z_X$ of the middle polariton. We normalize the population of the modes $n_\alpha(t)=N_\alpha(t)/\mathcal N_{\text{ss}}$ with the total number of excitations in the steady-state $\mathcal N_{\text{ss}}=\sum_\alpha N_\alpha(t\rightarrow \infty).$ The system is driven on resonance with the middle polariton, $\omega_p=\omega_X.$ 

We start with $J/\Omega = 1,$ in Fig.~\ref{fig:D1}(a), corresponding to a balanced hybridization where the steady state is formed by $50\%$ left cavity photon and $50\%$ exciton. Only at very small times, the right cavity occupation is comparable to the left cavity and exciton ones. Interestingly,  although the exciton couples only indirectly to the left cavity photons through the right cavity, the population of $n_R(t)$ remains strongly suppressed throughout the dynamics.

As the tunneling increases to $J/\Omega=2$ [[Fig.~\ref{fig:D1}(b)],  the excitonic component increases at the expense of the left-cavity photon component. Here, the degree of mixing gives $Z_L^{(MP)}=0.2$ and $Z_X^{(MP)}=0.8$ for the steady-state as indicated by the blue and black dashed lines and the long-time populations. The right-cavity photon contribution remains negligible, demonstrating the robustness of the intercavity character.

In the strong tunneling regime ($J/\Omega = 4$) [Fig.~\ref{fig:D1}(c)], the system evolves towards a state with dominant excitonic character while preserving its delocalized photonic component.  Notably, this crossover to a more matter-like polariton does not compromise the nonlocal polaritonic structure, a defining feature of intercavity polaritons.   These results show that coherent pumping of the middle polariton enables controlled tuning of the light–matter composition without sacrificing its intercavity nature.

Across the explored range of  $J/\Omega$  the dynamical formation of the intercavity polariton yields  steady-state populations in excellent agreement with the Hopfield model.  Furthermore, although photon tunneling between the cavities is essential to establish hybridization, the population of the right-cavity photon remains strongly suppressed throughout the dynamics, confirming the emergence of an effectively intercavity polariton.

\begin{figure}[h!]
    \centering
\includegraphics[width=0.993\columnwidth]{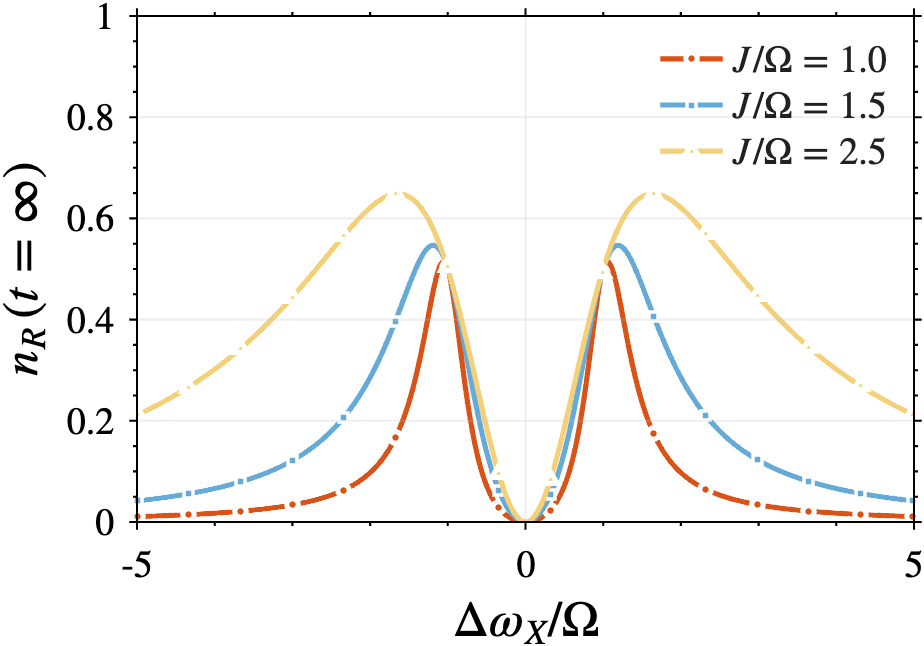}
    \caption{\textbf{Transparency window in the steady-state response.} 
    Steady-state population of the right cavity $n_R(t=\infty)$ as a function of the exciton detuning $\Delta\omega_X/\Omega$ for different tunneling-to-coupling ratios $J/\Omega$. A characteristic transparency window emerges around zero detuning, reflecting destructive interference between polariton pathways and the formation of a dark state.}
    \label{fig:D2}
\end{figure}

We now turn our attention to the study of the robustness of the intercavity polaritons. First, we analyze their formation as a function of the cavity detuning $\Delta_X=\omega_X-\omega_{p}$.  Figure.~\ref{fig:D2} shows the  steady-state population of the right cavity $n_R(t=\infty)$ for several values of the ratio $J/\Omega$.

We find that the  steady-state response exhibits a characteristic {\it transparency window} when the system is driven near resonance with the excitonic transition.  This feature is reminiscent of electromagnetically induced transparency (EIT) in quantum optical systems, where destructive interference suppresses the population of an intermediate state and signals the formation of a dark polariton. In the present case, the {\it transparency window} marks the emergence of an  intercavity polariton with its photonic and excitonic components spatially segregated.

\begin{figure}[t]
    \centering
    \includegraphics[width=0.993\columnwidth]{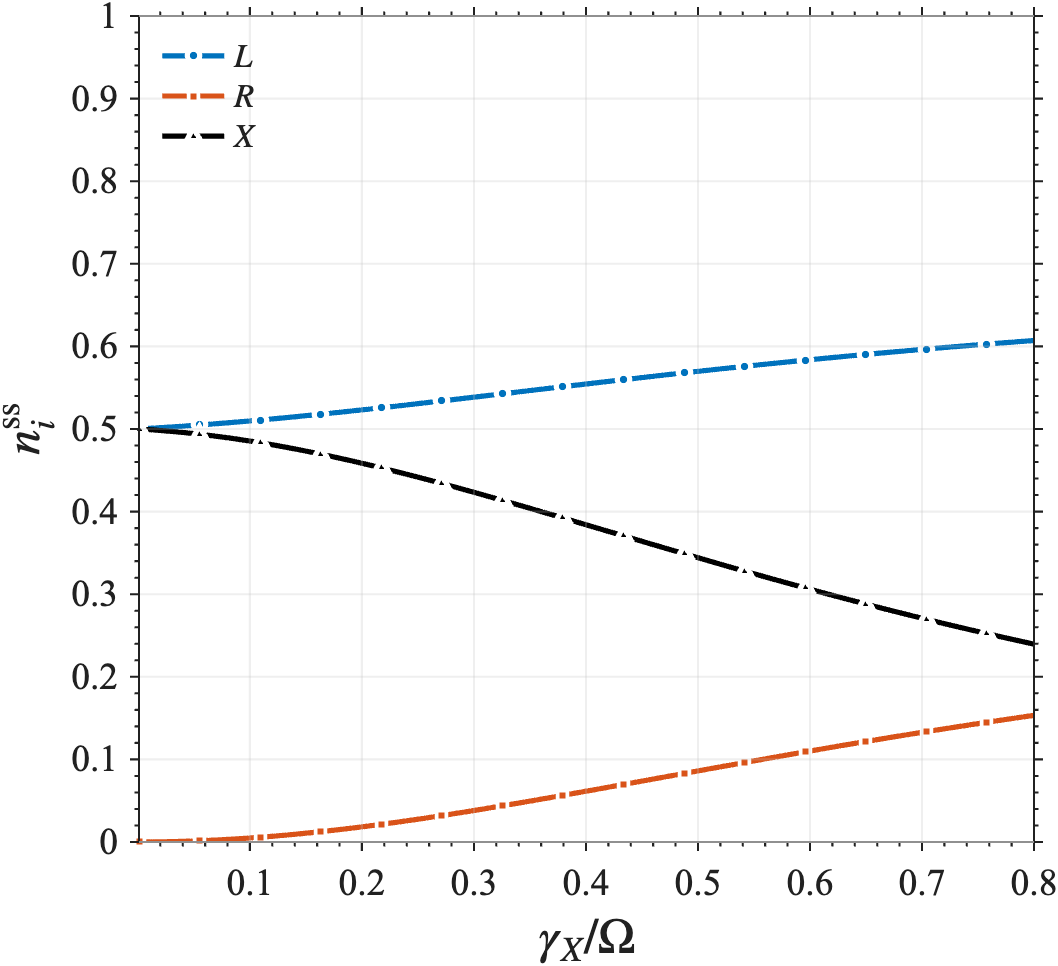}
    \caption{\textbf{Breakdown of intercavity polaritons due to exciton damping.} 
    Steady-state populations $n_i^{\mathrm{ss}}$ of the left cavity ($L$), right cavity ($R$), and exciton ($X$) components as a function of the normalized exciton decay rate $\gamma_X/\Omega$. Increasing $\gamma_X$ progressively suppresses the excitonic population and alters the balance of light–matter hybridization, ultimately destroying the collective polariton state.}
    \label{fig:D3}
\end{figure}

We next study the effects of a finite exciton lifetime. Figure~\ref{fig:D3}, shows the steady-state populations as the exciton damping rate $\gamma_X$ is increased. As $\gamma_X$ grows, the excitonic fraction $n_X^{\mathrm{ss}}$ decreases monotonically, indicating that dissipation suppresses the formation of coherent light–matter superpositions. 

At the same time, the photonic components exhibit compensating behavior: the population in the left cavity increases, while that in the right cavity remains suppressed but gradually increases as the system becomes more photon-like.

This behavior illustrates how strong exciton damping effectively breaks hybridization that underpins the formation of intercavity polaritons. We remark that as long as we remain in the strong light-matter coupling regime: $\gamma_X/\Omega<0.1$ the intercavity polariton is robust. 

\begin{figure*}[t]
    \centering
\includegraphics[width=\textwidth]{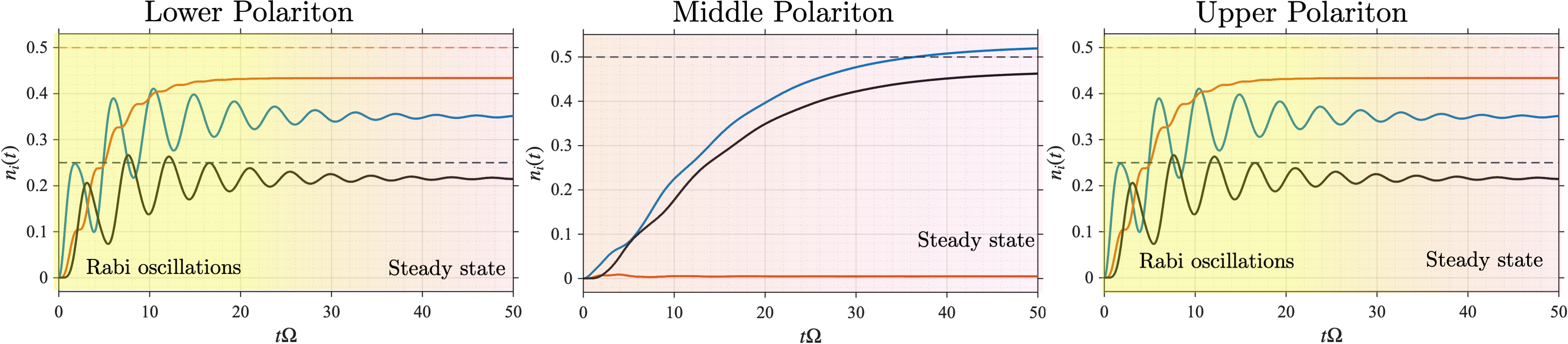}
    \caption{Time evolution of the populations $n_i(t)$ associated with the left-cavity photon (blue), right-cavity photon (orange), and exciton (black) components when the system is coherently driven at the eigenenergy of the (left) lower, (middle) middle, and (right) upper polariton branches. The lower and upper polariton dynamics exhibit pronounced Rabi oscillations before reaching their steady state, reflecting coherent light–matter exchange. In contrast, the middle polariton shows a smooth, non-oscillatory rise toward the steady state, a direct consequence of its purely intercavity character, which suppresses Rabi cycling. We take $J/\Omega=1$ and $\gamma_R/\Omega=0.5$ and $\gamma_L/\Omega=0.1$ }
    \label{fig:dynamics_polaritons}
\end{figure*}

An instructive way to visualize the dynamical formation of polaritons in the coupled-cavity system is to follow the time evolution of the populations associated with the three hybrid modes when the system is coherently driven at their respective eigenenergies. Figure~\ref{fig:dynamics_polaritons} shows the temporal dynamics of the left-cavity photon, right-cavity photon, and exciton populations when the pump frequency is tuned to the lower, middle, and upper polariton branches. The resulting behavior reveals markedly different routes toward quasiparticle formation, highlighting how the nature of each branch governs its transient evolution. In Figure~\ref{fig:dynamics_polaritons} we consider $\omega_X=\omega_R$ $J/\Omega=1$, $\gamma_L/\Omega=0.1$ and $\gamma_R/\Omega=0.5.$ 
We begin by considering a drive resonant with the lower polariton in Fig.~\ref{fig:dynamics_polaritons}(a). The populations of the left-cavity photons and excitons display pronounced oscillations, while the right-cavity photon population evolves more monotonically toward its steady state. Notably, the Rabi oscillations occur between the spatially segregated components of the polariton, with the left-cavity photon and exciton populations oscillating out of phase by approximately $\pi/2.$ These oscillations occur with a steady increase of the right-cavity photon occupation. Finally, the Rabi oscillations are attenuated and the stationary state is reached.

A similar behavior is observed for resonant driving of the upper polariton [Fig.~\ref{fig:dynamics_polaritons}(c)], reflecting the symmetry between the lower and upper branches for $\Delta_R=0$.

In contrast, driving the system resonantly at the middle polariton energy leads to qualitatively different dynamics, as shown in Figure~\ref{fig:dynamics_polaritons}(b). In this case, the population grows monotonically towards its steady-state value, with no pronounced oscillatory behavior. This absence of Rabi oscillations originates from the purely intercavity character of the middle polariton: its photonic and excitonic components reside in different cavities and are coupled only indirectly. As a result, the coherent  tunneling exchange that drives the oscillations in the lower and upper branches is suppressed, and the system instead evolves smoothly into a steady-state intercavity polariton. This behavior underscores the distinct dynamical nature of the middle branch and further illustrates how spatial separation of polariton components fundamentally alters their formation dynamics.

\section{Condensation and collective excitations of intercavity polaritons}
\label{{intercavityphonons}}

As demonstrated above, on resonance, the middle polariton (MP) emerges as a \emph{purely intercavity} quasiparticle.
At the same time, the MP retains a sizable excitonic fraction and, therefore, inherits nonlinear interactions, making it a promising candidate for 
macroscopic occupation and condensation under resonant driving.  
The coexistence of delocalized polaritonic coherence and matter-induced nonlinearities distinguishes 
the middle polariton from conventional lower polaritons and motivates a detailed study of its 
condensation and collective excitations.

To describe the formation of a condensate in the middle polariton branch, we start from the 
 microscopic Hamiltonian
\begin{gather}
\hat H_0
=
\sum_{\mathbf p}
\hat \Psi^\dagger_{\mathbf p}
\cdot
\!\!
\begin{pmatrix}
\omega_L(\mathbf p) & -J & 0 \\
-J & \omega_R(\mathbf p) & \Omega \\
0 & \Omega & \omega_X(\mathbf p)
\end{pmatrix}
\!\!\cdot
\hat \Psi_\mathbf p
\label{eq:H0_matrix}
\end{gather}
with $\hat \Psi^\dagger_{\mathbf p}=[\hat a^\dagger_{L}(\mathbf p),\hat a^\dagger_{R}(\mathbf p),\hat x^\dagger(\mathbf p)].$
Here, $\hat a_{L/R}(\mathbf p)$ annihilate photons in the left and right cavities, while 
$\hat x(\mathbf p)$ annihilates an exciton in the right cavity with momentum $\mathbf p$  and  energy $\omega_\alpha(\mathbf p) = \omega_\alpha^0 + \frac{|\mathbf p|^2}{2m_\alpha}$
where $\omega_\alpha^0$ denote the cavity and exciton resonance energies at normal incidence, and 
$m_\alpha$ are the corresponding effective masses.  
Because $m_L,m_R \ll m_X$, the exciton dispersion is much flatter than the photonic ones.  
The parameter $J$ denotes photon tunneling between cavities, and $\Omega$ is the Rabi coupling between the
right-cavity photon and the exciton.

In the absence of interactions, diagonalizing $\hat H_0$ yields three polariton branches.  We emphasize that even under resonance condition $\omega_L^0 = \omega_X^0,$ the middle polariton decouples from the right-cavity photon, strictly only at $\mathbf p=0.$ For finite momentum, the middle polariton takes the form
\begin{equation}
\hat b_\mathbf p
= \sqrt{Z_L(\mathbf p)}\,\hat a_L(\mathbf p)+\sqrt{Z_R(\mathbf p)}\,\hat a_R(\mathbf p) + \sqrt{Z_X(\mathbf p)}\,\hat x(\mathbf p),
\end{equation}
that is, a superposition of the three bare modes, with a contribution of the right-cavity photon to the middle polariton. 

At zero momentum, however, the middle polariton reduces to
\begin{equation}
\hat b_{\mathbf p=0}
= \sqrt{Z_L(\mathbf 0)}\,\hat a_L(\mathbf 0)+ \sqrt{Z_X(\mathbf 0)}\,\hat x(\mathbf 0),
\end{equation}
with $Z_R(\mathbf p=0)=0,$  corresponding to an ideal intercavity polariton.

Although pure intercavity polaritons are realized only at $\mathbf p=0,$ in Ref.~\cite{sanchez2025protecting}, it has been shown that the middle polariton retains its  intercavity nature for finite $\mathbf p.$  Here, we study the character of the collective excitations, that is, its composition in terms of an intercavity polariton or a full mixture of the three bare states.

We now include exciton–exciton interactions in the right cavity. Since the middle polariton  inherits a significant fraction of its matter character from the excitonic field, these nonlinearities play a crucial role in determining the collective dynamics near condensation. The interacting Hamiltonian reads
\begin{equation}
    \hat{H} = \hat{H}_0 + \frac{g_X}{2} \int d^2r\, \hat{x}^\dagger(\mathbf{r}) \hat{x}^\dagger(\mathbf{r}) \hat{x}(\mathbf{r}) \hat{x}(\mathbf{r}),
\end{equation}
where $g_X$ denotes the strength of the exciton–exciton interaction. This term introduces an effective nonlinearity into the middle polariton branch, enabling the emergence of a macroscopic coherent state and collective excitations beyond the single-particle picture. 

In the following, we project this interaction onto the middle polariton branch to obtain an effective low-energy description.

\subsection{Intercavity phonons}
We study the condensation of the middle polariton at $\mathbf{p}=0$ under coherent driving. In this setting, the external drive explicitly breaks the $U(1)$ symmetry; accordingly, we use the term condensation to refer to a macroscopically occupied driven–dissipative steady state within a mean-field description, rather than to an equilibrium phase transition.

As established in the dynamical analysis above, the MP state can be selectively populated. We therefore treat the $\mathbf p=0$ mode within a mean-field approximation replacing the field $\hat b_\mathbf 0\sim\hat b^\dagger_\mathbf 0\sim\sqrt{n_0},$ with $n_0$, i.e., the condensate density. The eigenvalues and eigenvectors are obtained from the Bogoliubov matrix, 
\begin{widetext}    
\begin{gather}
\mathcal L(\mathbf p)
=
\begin{pmatrix}
\epsilon_{\mathbf p} - \omega_p + 2n_0 g_{\rm MP}(\mathbf p)-i\gamma_{\text{MP}} & n_0 g_{\rm MP}(\mathbf p) \\
-n_0 g_{\rm MP}(\mathbf p) & -(\epsilon_{\mathbf p} - \omega_p + 2n_0 g_{\rm MP}(\mathbf p))-i\gamma_{\text{MP}}
\end{pmatrix}.
\label{eq:HD_matrix}
\end{gather}
\end{widetext}
Here, $g_{\text{MP}}(\mathbf p)$ denotes the effective polariton--polariton interaction,
renormalized by the quasiparticle residue of the excitonic component of the middle
polariton. Since polaritons interact via their exciton fraction, one has
\begin{equation}
 g_{\text{MP}}(\mathbf p) = g_X\, Z_X(\mathbf p)\, Z_X(\mathbf 0),
\end{equation}
with $g_X$ the bare exciton--exciton interaction strength and $Z_X(\mathbf p)$
the excitonic Hopfield coefficient of the middle polariton at momentum $\mathbf p$.
The quantity $\epsilon_{\mathbf p}$ denotes the single-particle dispersion of the
middle polariton branch.

Assuming a momentum-independent interaction, $g_{\mathrm{MP}}(\mathbf{p}) \simeq g_{\mathrm{MP}}(0) \equiv g_{\mathrm{MP}}$, an approximation that is justified a posteriori by the weak momentum dependence of the Hopfield coefficients in the relevant regime, and fixing $\omega_p = n_0 g_X Z_X^2(0)$, we obtain the Bogoliubov excitation spectrum
\begin{equation}
 E_{\mathbf p}
 =
 \pm \sqrt{\epsilon_{\mathbf p}
 \bigl(\epsilon_{\mathbf p} + 2 g_{\text{MP}} n_0\bigr)}-i\gamma_{\text{MP}}.
 \label{eq:Bogoliubov_MP}
\end{equation}
In this regime, damping introduces a small imaginary component to the excitation spectrum. We consider the weak-damping regime $g_{\text{MP}}n_0\gg \gamma_{\text{MP}},$ and we focus on the real part of the Bogoliubov spectrum. 

\begin{figure}[t]
    \centering
    \includegraphics[width=.5\textwidth]{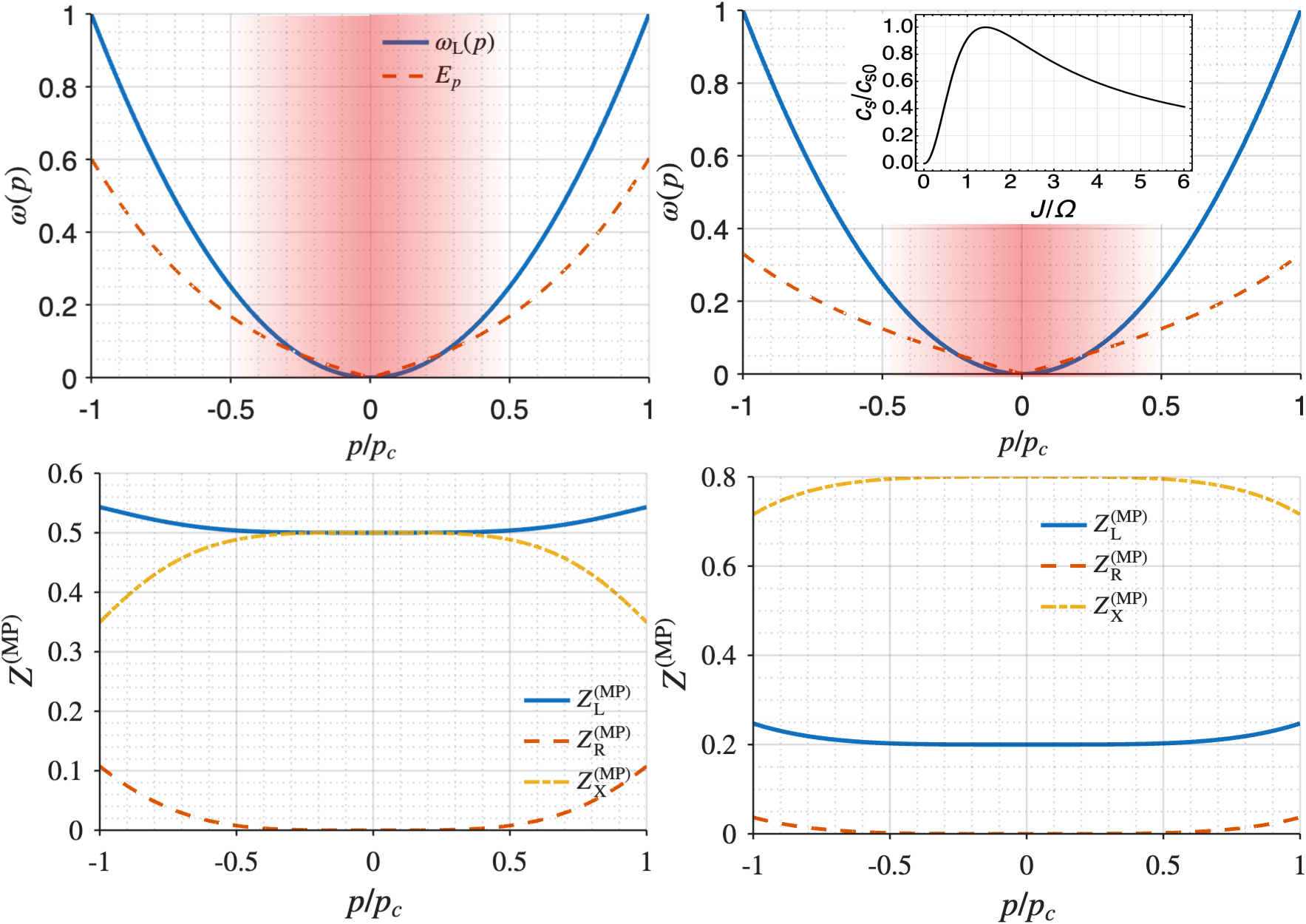}
    \caption{Bogoliubov dispersion of the middle polariton condensate (top) and quasiparticle residues (bottom) for $J/\Omega=1$ (left) and $J/\Omega=2.0$ (right). For the quasiparticle dispersion, the solid blue line shows the bare cavity photon dispersion $p^2/2m_c$, while the dashed red line shows the Bogoliubov excitation spectrum. In the bottom panel we show that the quasiparticle residue $Z_{R}^{\text{MP}}$ essentially vanishes for a large range of momentum. The inset shows the speed of sound as a function of $J.$ }
    \label{fig:dispersion}
\end{figure}

For low momenta, $\epsilon_{\mathbf p} \simeq |\mathbf p|^2/(2 m_{\text{MP}})$, this reduces to a
phonon-like dispersion $E_{\mathbf p} \simeq c_s |\mathbf p|$ with sound velocity
\begin{equation}
 c_s = \sqrt{\frac{g_{\text{MP}} n_0}{m_{\text{MP}}}},
 \label{eq:cs_general}
\end{equation}
here $m_{\text{MP}}/m_L=1+\frac{J^2}{\Omega^2}.$ In our intercavity geometry, $g_{\text{MP}}$ is set by the excitonic fraction of the
middle polariton, while $m_{\text{MP}}$ is governed by its photonic component. Note that, owing to the intercavity character of the polariton, the speed of sound can be tuned locally: the interaction strength $g_X$ and the effective mass $m_L,$ enabling independent control of interactions and kinetic energy.  In Fig.~\ref{fig:dispersion} (inset) we show the non-monotonous dependence of the speed of sound $c_s$ as a function of $J,$ normalised with respect to its maximum $c_{s0}$ at $J/\Omega=\sqrt{2}.$  The sound velocity
can be  tuned by the ratio $J/\Omega$, while remaining insensitive to
the right-cavity detuning $\Delta_R$ within our model. 

A key result is that the phonon-like regime of the Bogoliubov dispersion is independent of the right-cavity detuning. The linear spectrum
holds whenever $
\frac{p^2}{2m_{\mathrm{MP}}} \ll 2 g_{\mathrm{MP}} n_0.$ 
In this limit, both  $m_{\mathrm{MP}}$ and $g_{\mathrm{MP}}$ are fixed by the left-cavity photon effective mass and by the excitonic residue of the middle polariton at $p=0$,  and are therefore insensitive to the right-cavity detuning within our model.

In Fig.~\ref{fig:dispersion}, we show the Bogoliubov dispersion of the middle polariton (top row) together with the corresponding Hopfield coefficients (bottom row) for two different values of $J/\Omega$. The ratio $J/\Omega$ controls the slope of the linear (phonon-like) regime at low momenta, as seen by comparing $J/\Omega=1$ [Fig.~\ref{fig:dispersion}(a)] and $J/\Omega=2$ [Fig.~\ref{fig:dispersion}(b)]. For small momenta, $p/p_c \ll 1$, with $p_c=\sqrt{2m_{\text{MP}}\Omega}$, the dispersion is linear, reflecting the sound-like behavior of the condensate. As $p/p_c$ increases, the spectrum crosses over to a quadratic dispersion and asymptotically approaches the bare polariton dispersion. This crossover from the phonon-like to the single-particle regime is highlighted by the shaded region in the figure.

To further characterise the nature of the Bogoliubov excitations,  we show the quasiparticle residues~\ref{fig:dispersion} (bottom row). In the regime where the dispersion is linear we find that the polariton remains purely intercavity, that is, in this regime the collective excitations of the medium are hybrid light-matter phonon-like quasiparticles with their components spatially segregated.

Importantly, despite the collective nature of the excitations, the right-cavity photonic component remains strongly suppressed, demonstrating that the Bogoliubov quasiparticles inherit the intercavity character of the underlying polariton.

\section{Experimental Perspective}
\label{experiments}
Intercavity polaritons have already been experimentally demonstrated in organic microcavity systems, where
Frenkel excitons confined to one cavity hybridize with photons in a second, spatially separated cavity. These
experiments confirm the existence of a middle polariton branch with a purely intercavity character, consistent
with the theoretical structure analyzed here. However, Frenkel polaritons exhibit only weak exciton-exciton
interactions due to the highly localized nature of the organic excitons, making them ideal for realizing long-lived,
dark-state-like hybrid modes but less suitable for exploring strongly interacting regimes or condensation
phenomena driven by nonlinearities. To achieve a bona fide polariton Bose-Einstein condensate in an intercavity
geometry, and to access regimes where collective excitations and Bogoliubov physics become observable,
inorganic semiconductor microcavities (such as GaAs, ZnO, or transition-metal dichalcogenide excitons) would 
likely be required. These systems provide significantly stronger exciton-exciton interactions and enhanced 
nonlinearities, while retaining the tunability and coherence properties that enable the formation of intercavity 
polaritons. Thus, the combination of an intercavity architecture with strongly interacting inorganic polaritons 
offers a promising route toward realizing an intercavity polariton BEC.

Finally, the experimental observation of tunable dispersion without sacrificing photonic weight provides
a clear motivation for pursuing condensation in the intercavity middle polariton branch. In contrast to
conventional two-level polaritons, where flattening the dispersion typically requires reducing the light-matter
mixing, the three-level architecture enables a robust effective mass while retaining sizeable photon content.
This opens the door to engineering collective phenomena, including polariton condensation, superfluidity,
and nonlinear response, in a regime where the quasiparticle mass is fixed but the interaction scale and
Bogoliubov spectrum are highly tunable. The demonstrated experimental control over detuning, lifetime,
and spatial composition points to promising pathways for realizing long-lived, interaction-enhanced
intercavity polariton condensates consistent with the theoretical framework developed in this work.

\section{Conclusions}
\label{conclusions}
We have presented a comprehensive theoretical study of the dynamical formation of intercavity polaritons in strongly coupled cavity systems. By solving the dynamical equations under coherent resonant driving, we have shown that the middle polariton branch enables the emergence of a hybrid quasiparticle with uniquely spatially segregated photonic and excitonic components. The Hopfield analysis demonstrates that at resonance this mode suppresses the right-cavity photon field entirely, yielding a polariton composed solely of left-cavity photons and right-cavity excitons.

Time-resolved simulations reveal strikingly different dynamical behavior among the three polariton branches. While the lower and upper polaritons exhibit characteristic Rabi oscillations due to coherent light–matter exchange, the middle polariton evolves smoothly and monotonically toward its steady state, reflecting its non-local structure. This fundamental distinction provides a clear dynamical signature of the intercavity polariton and underscores how spatial separation qualitatively modifies quasiparticle formation.

We further demonstrate that intercavity polaritons are both robust and highly tunable. Detuning and photon tunneling enable control over light–matter hybridization and give rise to transparency windows, while exciton lifetime plays a key role in preserving coherence. These results establish dynamical formation as a powerful probe of intercavity polaritons and a route toward engineering their properties beyond steady-state descriptions.

More broadly, our work shows that the intercavity architecture enables independent control over dispersion and interactions, opening a pathway toward realizing interacting, driven–dissipative polariton condensates with spatially structured light–matter composition. This provides a versatile platform for exploring collective quantum phenomena in multimode cavity systems.

\section{Acknowledgments}
A.C-G. acknowledges financial support from UNAM DGAPA PAPIIT Grant No. IA101325, Project CONAHCYT No. CBF2023-2024-1765 and PIIF25. G.P. acknowledges financial support from UNAM DGAPA PAPIIT Grant No. IN104325 and project PIIF 2025. H.A.L-G. acknowledges financial support from UNAM DGAPA PAPIIT Grant No. IN106725 and Project SECIHTI No. CBF-2025-I-2241.
\appendix
\begin{widetext}
\section{Dynamics}

In this Appendix we provide details on the analysis of the out-of-equilibrium dynamics. From the spinor
\begin{gather}
\hat \Psi = 
\big[\hat a_L(t),\;\hat a_R(t),\;\hat x(t)\big]^T,
\end{gather}
we write the Heisenberg equations of motion in the rotating frame of the pump frequency $\omega_p$ as
\begin{gather} \label{eq:3_Heisenberg_motion_rot_frame}
  i\frac{\partial \hat \Psi(t)}{\partial t}=\hat H \cdot \hat \Psi(t)+\mathbf F,
\end{gather}
with the driving vector $\mathbf F=[F,\,0,\,0]^T$, and effective Hamiltonian
\begin{gather} \label{eq:effective_hamiltonian}
\hat H=
\begin{bmatrix} 
 -\Delta_L & -J & 0 \\
 -J & -\Delta_R & \Omega \\
 0 & \Omega & -\Delta_X
\end{bmatrix},    
\end{gather}
where $\Delta_\alpha=\omega_p-\omega_\alpha+i\gamma_\alpha$ as mentioned in the main text.

The explicit form of the equations of motion are

\begin{gather}
 i\frac{\partial}{\partial t} \hat a_L(t) =   -\Delta_L \hat a_L(t) - J \hat a_R (t) + F   ,\\
\nonumber
i\frac{\partial}{\partial t}\hat a _R(t) =   -\Delta_R  \hat a_R (t) - J \hat a_L (t)+ \Omega \hat x  (t) ,  \\
\nonumber  
i\frac{\partial}{\partial t} \hat x (t) =   -\Delta_X\hat x (t)   + \Omega \hat a_R (t)   .\\
\nonumber
\end{gather}

Now, for the steady-state we have 
\begin{gather}
    i\frac{\partial \hat \Psi(t)}{\partial t}=0.
\end{gather}
Taking expectation values, we define the steady-state classical fields
\begin{equation}
    \Psi_{ss} \equiv \langle \hat{\Psi}(t\to\infty)\rangle,
\end{equation}
which satisfy the same equations of motion due to their linearity. The steady-state solution is then given by
\begin{gather}
\hat\Psi_{ss}=-\hat H^{-1}\cdot \mathbf F=\\ \nonumber   = \frac{1}{J^2 \Delta _X - \Delta_X^2 \Delta _R + \Omega^2 \Delta _X} \begin{bmatrix}
 \Delta_R \Delta_X - \Omega^2 & 
 - \Delta_X J & 
- \Omega J
\\
 - \Delta _X J & 
\Delta _X^2 & 
 \Delta _X \Omega\\
- \Omega J & 
 \Delta _X \Omega & 
\Delta _R \Delta_X - J^2
\end{bmatrix}\cdot \mathbf F 
=-\begin{bmatrix}
\displaystyle \frac{F \left( \Delta_R \Delta_X - \Omega^2 \right)}{J^2 \Delta _X - \Delta_X^2 \Delta _R + \Omega^2 \Delta _X} 
\\\frac{- \Delta _X J\ F}{J^2 \Delta _X - \Delta_X^2 \Delta _R + \Omega^2 \Delta _X} \\
 \frac{- \Omega J \ F}{J^2 \Delta _X - \Delta_X^2 \Delta _R + \Omega^2 \Delta _X} 
\end{bmatrix}.
\end{gather}
For $\Delta_X$ small, we find that the contribution to the steady-state are governed by
\begin{gather}
\Psi_{ss}=\frac{\Omega F}{\sqrt{J^2+\Omega^2}\Delta_X}\begin{bmatrix}
\frac{   \Omega }{\sqrt{J^2  + \Omega^2 }} \\
0 
\\
\frac{   J}{\sqrt{J^2  + \Omega^2 }} 
\end{bmatrix}=\sqrt{\Psi_{0}}\begin{bmatrix}
\frac{   \Omega }{\sqrt{J^2  + \Omega^2 }} \\
0 
\\
\frac{   J}{\sqrt{J^2  + \Omega^2 }} 
\end{bmatrix} =\sqrt{\Psi_0}\left[\cos\theta|L\rangle+\cos\theta|X\rangle \right]
\end{gather}
Thus, we identify as the total number of excitations as $N_{\text{tot}}^{(0)}=|\Psi_0|^2.$  The structure of the steady-state precisely recovers the structure of the intercavity polaritons with a total number of polaritons $N_{\text{tot}}^{(0)}.$
\end{widetext}
\bibliography{F_Polariton}
\end{document}